## IN PRESS

# Single Molecule FRET Reveals Pore Size and Opening Mechanism of Msc

Yong Wang (University of Illinois at Urbana-Champaign), Yanxin Liu (University of Illinois at Urbana-Champaign), Hannah DeBerg (University of Illinois at Urbana-Champagin), Takeshi Nomura (Victor Chang Cardiac Research Institute), Melinda Hoffman (University of Illinois at Urbana-Champaign), Paul Rohde (Victor Chang Cardiac Research Institute), Klaus Schulten (University of Illinois at Urbana-Champaign), Boris Martinac (Victor Chang Cardiac Research Institute), and Paul Selvin (University of Illinois, Urbana-Champaign)

**Abstract:**

The mechanosensitive channel of large conductance, which serves as a model system for mechanosensitive channels, has previously been crystallized in the closed form, but not in the open form. Ensemble measurements and electrophysiological sieving experiments show that the open-diameter of the channel pore is >25Å, but the exact size and whether the conformational change follows a helix-tilt or barrel-stave model are unclear. Here we report measurements of the distance changes on liposome-reconstituted MscL transmembrane α-helices, using a "virtual sorting" single-molecule fluorescence energy transfer. We observed directly that the channel opens via the helix-tilt model and the open pore reaches 2.8 nm in diameter. In addition, based on the measurements, we developed a molecular dynamics model of the channel structure in the open state which confirms our direct observations.

# Single Molecule FRET Reveals Pore Size and Opening Mechanism of MscL

Yong Wang[1,2], Yanxin Liu[1,2], Hannah A. DeBerg[1,2], Takeshi Nomura[3], Melinda Tonks Hoffman[1,2], Paul R. Rohde[3], Klaus Schulten[1,2], Boris Martinac[3,4], Paul R. Selvin[1,2,5,*]

[1]Department of Physics, [2]Center for the Physics of Living Cells, [5]Center for Biophysics and Computational Biology, University of Illinois at Urbana-Champaign, Urbana, IL, USA, [3]Molecular Cardiology and Biophysics Division, Victor Chang Cardiac Research Institute, [4]University of New South Wales, Sydney, Australia.

[*] To whom correspondence may be addressed:
    Paul R. Selvin
    Telephone: (217) 244-3371
    Fax: (217) 333-4898
    Email: selvin@illinois.edu

Running Title: smFRET study on the opening of MscL

ABSTRACT

The mechanosensitive channel of large conductance, which serves as a model system for mechanosensitive channels, has previously been crystallized in the closed form, but not in the open form. Ensemble measurements and electrophysiological sieving experiments show that the open-diameter of the channel pore is >25Å, but the exact size and whether the conformational change follows a helix-tilt or barrel-stave model are unclear. Here we report measurements of the distance changes on liposome-reconstituted MscL transmembrane α-helices, using a "virtual sorting" single-molecule fluorescence energy transfer. We observed directly that the channel opens via the helix-tilt model and the open pore reaches 2.8 nm in diameter. In addition, based on the measurements, we developed a molecular dynamics model of the channel structure in the open state which confirms our direct observations.

**Introduction**

Mechanosensitive (MS) channels are essential in both eukaryotes and prokaryotes (Haswell *et al*, 2011; Árnadóttir & Chalfie, 2010; Perozo, 2006). In eukaryotes, they are involved in diverse processes such as embryonic development, touch, pain, hearing, lung growth, and muscle homeostasis (Chalfie, 2009; Hamill & Martinac, 2001; Árnadóttir & Chalfie, 2010). In bacteria, they are "safety valves", opening their pores to release the pressure to protect cells from hypo-osmotic shock (Booth & Blount, 2012). The rise in antibiotic resistance, and the crucial role MS channels play in bacterial adaptation, makes it important to understand the MS channels as potentially new drug targets (Booth & Blount, 2012).

When high pressure (~ 10 mN/m) causes the bacterial mechanosensitive channel of large conductance (MscL) to open, it forms a large, nonselective pore with a very high conductance (~ 3 nS) that is permeable to various ions and small organic osmolytes. In 1998, MscL from *Mycobacterium tuberculosis* in the closed state was crystallized by Rees and co-workers (Chang *et al*, 1998). They showed that MscL is a pentamer made up of five identical subunits (Figure 1a-b). Each subunit consists of one cytoplasmic α-helix (the CP domain) and two trans-membrane α-helices (the TM1 and TM2 helices), which extend through the cell membrane and are joined by a periplasmic loop (Figure 1b). TM1 and TM2 are primarily responsible for gating; it has been shown that complete deletion of the CP domain does not change the gating parameters substantially (Anishkin *et al*, 2003).

Despite this progress, the open form of MscL has not been crystallized. This leaves two questions unanswered: what is the exact size of the open pore of MscL, and how does the channel open? Several techniques, e.g., permeation of organic ions (Cruickshank *et al*, 1997), Electron Paramagnetic Resonance (EPR) (Perozo *et al*, 2002b, 2002a) and ensemble Fluorescence Resonance Energy Transfer (FRET) (Corry *et al*, 2005b, 2010) have attempted to measure the pore size. However, systematic errors likely result in an overestimation of (Cruickshank *et al*, 1997), an underestimation of (Corry *et al*, 2005b, 2010), or an insensitivity to the requisite distances (Perozo *et al*, 2002a). For example, EPR was only able to establish that the open pore is > 25Å (11). Ensemble FRET, which yielded some insightful results, is potentially sensitive to larger distances (~ 80-100 Å) (Roy *et al*, 2008). However, due to multiple labeling, problems with protein clustering, and the need for Monte-Carlo simulations to extract distance information, there was much variability and uncertainty in the results (Corry *et al*, 2005b, 2010, 2005a).

Another important question is how the MscL channels open, i.e. how the helices rearrange upon channel activation (i.e., from the closed state to the open state). Currently, there exist two predominant models for the opening of MscL: the barrel-stave model and the helix-tilt model (Figure 2) (Perozo, 2006). The barrel-stave model involves motion of the transmembrane helix 1 (TM1) with the transmembrane helix 2 (TM2) remaining stationary; the open-pore is lined by both TM1 and TM2 and the helices are fairly vertical (where the membrane is horizontal). This model derives primarily from the number of transmembrane helices and the large size of the open pore. In contrast, the helix-tilt model, which has been proposed more recently (Sukharev *et al*, 2001a, 2001b; Betanzos *et al*, 2002), involves motion of TM1 and TM2, with both swinging away from the pore upon channel opening and both helices tilting toward the plane of membrane. Recent evidence from cysteine-crosslinking experiments, EPR experiments, and ensemble FRET experiments, argue in favor of the helix-tilt model (Betanzos *et al*, 2002; Perozo *et al*, 2002a; Corry *et al*, 2010).

In the present work, we focused on the transmembrane helices involved in the opening of MscL from *Escherichia coli* (EcoMscL), using a single-molecule fluorescence resonance energy transfer (smFRET). MscL channels were reconstituted in liposomes during smFRET measurements and thus the channels were in their quasi-native environment. In addition, although MscL is a pentamer, we utilized photobleaching to virtually sort out the population of molecules with a single donor and a single acceptor, allowing us to make accurate smFRET measurements. It is the first time that smFRET has been applied to liposome-reconstituted membrane proteins with more than three monomers. We measured movement of three residues on TM1 (M42C, A27C, and I25C; Figure 1c) and three residues on TM2 (Y75C, Q80C and V82C; Figure 1c), from which we determined not only the translational movements but also the tilting of each helix. We observed the tilting of the helices in a model-free fashion, arguing strongly in favor of the helix-tilt model. In addition, from the movement of the residue (I25C) right at the gating region, we determined directly that the open pore reaches 2.8 nm in diameter. Lastly, we developed a molecular dynamics model of the channel structure in the open state based on the smFRET results, while using the crystal structure of the protein in the closed state as a reference. The model of the open structure satisfies all the distance constraints measured from smFRET experiments. The developed open structure confirmed that the pore size of the fully open channel is 2.8 nm in diameter, achieved via the helix-tilt opening model.

**Results**

**FRET efficiencies.** Purified MscL mutants (Supplementary Figure 1) were labeled with Alexa Fluor 488 (AF488) and Alexa Fluor 568 (AF568) and reconstituted into ~ 50 nm liposomes made of 1-palmitoyl-2-oleoyl-sn-glycero-3-phosphocholine (POPC) with 2% 1,2-dioleoyl-sn-glycero-3-phosphoethanolamine-N-biotinyl (BPE) (Figure 1e). The liposomes were then immobilized on a glass coverslip, via biotin-avidin interaction, for smFRET measurements (Figure 1e). To access the open state of the channels, 1-oleoyl-2-hydroxy-sn-glycero-3-phosphocholine or lysophosphatidylcholine (LPC) of 25% molar ratio was added (Perozo *et al*, 2002b, 2002a; Corry *et al*, 2005b, 2010) (Figure 1f) and incubated for > 10 minutes before immobilization. Just before performing smFRET experiments, the fluorescence spectra of the samples (± LPC) were recorded with excitation at 488 nm to confirm that the channels open up with LPC by observing the shift in the FRET peaks. (The channel activity is also determined by observing the opening of the channels upon application of negative pressure (suction) to the patch pipette. The labeled proteins for patch-clamp experiments are from a different aliquot, although the same batch, of the labeled proteins for smFRET experiments.) We emphasize that, although smFRET has been applied to study the conformational changes of channels and transporters (Zhao *et al*, 2010, 2011; Akyuz *et al*, 2013; Choi *et al*, 2010), to our knowledge, it is the first time that smFRET has been used with channels reconstituted to liposomes.

Via smFRET measurements, we observed fluorescence intensity traces with one or two photobleaching steps (Supplementary Figure 2a-b). This is the expected result because MscL is a homopentamer and the labeling of fluorophores is stochastic. The number of photobleaching steps tells the number of fluorophores attached to a channel. Only the traces showing a single photobleaching step in both the donor and acceptor channels, ensuring that only a single donor and/or acceptor fluorophore, were included in the analysis (Supplementary Figure 2a). Donors were, in most cases, photobleached first, resulting in simultaneous dropping of the fluorescence intensities in both donor and acceptor channels (Supplementary Figure 2a-b). Subtraction of the intensities before and after photobleaching gives the intensities of donor ($I_D$) and acceptor ($I_A$), which are used for the calculation of FRET efficiency. Note that the intensities, $I_D$ and $I_A$, automatically remove the direct excitation of acceptor (i.e. the leakage of acceptor emission in the donor channel). However, the leakage of donor emission in the acceptor channel is still present. To measure this leakage, MscL channels were labeled with donors-only and the leakage coefficient ($\ell$) was measured experimentally: $\ell = I_D^A/I_D^D \approx 0.09$, where $I_D^A$ is the intensity of donor emission in the acceptor channel and $I_D^D$ is the intensity of donor emission in the donor channel. Furthermore, to determine the actual FRET efficiency, another instrumental correction was made through the correction factor $\gamma$, which accounts for the differences in quantum yield and detection efficiency between the donor and the acceptor. It was calculated as the ratio of change in the acceptor intensity, $\Delta I_A$, to the change in the donor intensity, $\Delta I_D$, upon acceptor photobleaching: from the traces where the acceptor photobleached first, we estimated the value $\gamma = \Delta I_A/\Delta I_D \approx 0.89 \pm 0.06$ (Supplementary Figure 2c).

We analyzed a few hundred traces (varying between 134 and 577 traces) with single photobleaching steps in the absence and presence of LPC for each mutant (Figure 1c and Figure 3). Here we focus on the mutant M42C for the sake of illustration. For the single photobleaching steps of M42C, 428 and 577 traces, in the absence and presence of LPC, respectively, were analyzed. The corrected FRET efficiencies were calculated and their distribution was then plotted and fitted with Gaussians via maximum likelihood estimates, shown in Figure 3a-b, while the number of Gaussians was determined according to the corrected Akaike information criterion (AICc) and the Bayesian information criterion (BIC) (Supplementary Table I) (Schwarz, 1978; Akaike, 1974; Sugiura, 1978). In the absence of LPC, we observed three peaks at E = 0.1, 0.28 and 0.63, respectively (Figure 3a). In the presence of LPC, the third peak showing the highest FRET efficiency diminishes, leaving mainly two Gaussians (E = 0.1 and 0.23, Figure 3b). This transition (i.e. the highest peak decreases and the lowest peak increases) is more obvious when we plotted the difference between the normalized FRET distributions ($\sum P^X = 1$, where X = + for in the presence of LPC and X = − for in the absence of LPC) under the two conditions, as shown in Figure 3c: after adding 25% LPC, the peak at E ~ 0.6 went away but the fraction of the peak at E ~ 0.1 built up. Note that the highest peak at E ~ 0.6 does not completely disappear in the presence of 25% LPC, which is consistent with Ref. (Perozo *et al*, 2002b).

In the absence of LPC, the existence of three peaks, rather than two peaks, can be explained by considering the effect of tethering on the liposome. As the MscL channel is a homo-pentamer, we initially expected two distances between donor and acceptor in each state ($R_n$ and $R_f$ in Figure 5a-b) and thus two peaks for the distribution of FRET efficiency, assuming that all the channels are closed in the absence of LPC. However, this assumption is not necessarily true, especially in our situation where liposomes are immobilized and the proteins are responsive to membrane tension. It had been predicted by theories and observed in experiments that immobilization of liposomes (or vesicles) results in significant membrane tension and possibly rupture (Serro *et al*, 2012; Zhdanov *et al*, 2006; Chung *et al*, 2009). In our experiments, the membrane tension is expected to be high, ~ 30-40 $k_B T$, due to the strong interaction between BPE and the surface via biotin-neutravidin (Miyamoto & Kollman, 1993; Rico & Moy, 2007). With such strong interaction, giant unilamellar vesicles ruptured spontaneously, as has been observed experimentally (Chung *et al*, 2009). The consequence is that some of the MscL channels switch to the open conformation upon the immobilization of the liposomes. (However, the fraction of open channels might be different for different mutants even if the membrane tension is similar.) Therefore, the FRET histogram for the no-LPC sample includes a mixture of closed and open MscL channels. This hypothesis is supported by our results that show the number of open channels increases with the fraction of BPE (Figure 3d; see Supplementary Information for details). In addition, a simple estimation based on the crystal structure (Chang *et al*, 1998) and previously predicted open pore-size (Corry *et al*, 2010) gives that $R_n$ in the open state ($R_{no}$) is exactly the same as $R_f$ in the closed state ($R_{fc}$), indicating that it is very likely 192 that the middle peak is an overlap of two peaks corresponding to $R_{no}$ and $R_{fc}$. Furthermore, it has been
193 observed that ≤ 30% of MscL are hexamers, instead of pentamers, in detergents such as n-Dodecyl-β-D-
194 maltopyranoside (DDM), used in the current study. This would tend to "smear" the middle peaks of
195 FRET in the absence of LPC. Therefore to be consistent and accurate, we always use the highest FRET
196 efficiency for the calculation of distance changes. On the other hand, we did find that all mutants give $E_{fo}$
197 measurements compatible (i.e., within error) with the final model (except that M42C is slightly off), as
198 shown in Supplementary Figure 9.

199   FRET between neighboring MscL channels on the same liposome had been a problem in previous
200 ensemble FRET experiments. To decrease the likelihood of its happening, and to effectively solve the
201 problem, we applied two strategies. First, we used 5% labeled channels together with 95% unlabeled ones
202 for reconstitution in liposomes. As a result, we had 16x lower molar ratio of labeled proteins (pentamers)
203 to lipids than that in the ensemble FRET experiments: 1:4000 *vs.* 1:250 (Corry *et al*, 2010, 2005b), greatly
204 reducing the likelihood of inter-molecular FRET. In addition, only traces showing a single photobleaching
205 step in both donor and acceptor channels were included in analysis, which helps further removing the
206 FRET between neighboring MscL channels in the analysis.

207   Another note is that we used maximum-likelihood estimation (MLE) (In Jae, 2003) for peak fitting.
208 This method was chosen particularly because it does not require binning the data before fitting. Although
209 there are mathematical ways for selection of "good" bin sizes (Shimazaki & Shinomoto, 2007), the
210 selection of bin size is, in practice, subjective, and the peaks derived can be affected with different bin
211 sizes. After MLE fitting, we then bin the data and plot the histograms for the sake of presentation purpose.
212 How the data is binned does not change the fitting parameters.

214 **Measurement of Förster radius, $R_0$.** The Förster radius ($R_0$) for AF488 and AF568 is calculated by
215 means of $R_0 \propto (\kappa^2 \, Q_D)^{1/6}$ (Förster, 1948; Iqbal *et al*, 2008). Because $\kappa^2$ and $Q_D$, can be environmentally
216 sensitive, we measured the quantum yield and orientation factor for the fluorophores conjugated to each
217 and every channel mutant (Fery-Forgues & Lavabre, 1999; Lakowicz, 1999) (Figure 4). The quantum
218 yields of AF488 conjugated to various MscL mutants are summarized in Table I and Figure 4a, corrected
219 for polarization effects (Lakowicz, 1999; Fery-Forgues & Lavabre, 1999). It is noted that the fluorophores
220 used in the current study are mixtures of 5' and 6' isomers. However, it was expected that this will not
221 affect the results because 1) they have successfully been used in many smFRET studies (Granier *et al*,
222 2007; Jäger *et al*, 2006; Majumdar *et al*, 2007; Marras *et al*, 2002; Yin *et al*, 2005); 2) the chromophores
223 of the isomers are exactly the same while the only difference between the isomers lies in where the linker
224 of carbon-chain [$(CH_2)_5NHCO$] is attached; 3) we examined the molecular structures of the probe-isomers
225 and confirmed that the difference in molecular size is < 5% between isomers (see Supplementary Figure
226 7). The orientation factor $\kappa^2$, was determined by measuring the anisotropy of the conjugated fluorophores
227 (Table I, Figure 4b and Supplementary Figure 3). The anisotropy of both donor and acceptor for most
228 residues is < 0.2 and therefore $\kappa^2$, in fact, is close to 2/3 (Roy *et al*, 2008; Clegg, 1992; Andrews &
229 Demidov, 1999). Nonetheless, we calculated the maximum possible errors in $R_0$ due to anisotropic
230 orientation of the dyes (see Table I, Figure 4b and Supplementary Figure 3); the actual errors in $R_0$ should
231 be much smaller. Another source of error in $R_0$ lies in the measurements of $Q_D$, which were performed for
232 AF488-MscL in detergent (PBS + 1mM DDM), which was not exactly the same environment for
233 fluorophore-MscL conjugates in smFRET experiments (incorporated in liposomes in PBS), although the
234 buffer was kept the same. Furthermore, it is possible that the addition of LPC and the conformational
235 change of MscL changes $Q_D$ as well, resulting in additional errors in $R_0$ and in the distances calculated
236 below.

238 **Estimation of residue movements.** We measured the change of FRET efficiency of MscL before and
239 after channel activation using smFRET. For example, for M42C, the FRET efficiency changed from 0.63

240 (closed state) to 0.23 (open state). We also determined experimentally the Förster radii ($R_0 = 5.5^{+0.4}_{-0.3}$ nm
241 for M42C). This permitted us to estimate the change in the distance between donors and acceptors from
242 the closed to open states (Figure 5), $\Delta R_n = R_{no} - R_{nc} = R_0 (E_{no}^{-1} - 1)^{-1} - R_0 (E_{nc}^{-1} - 1)^{-1}$ ($\approx$ 1.7 nm for M42C).
243 We emphasize that some of the distances between fluorophores ($R_{no}$ and $R_{nc}$ in Supplementary Table II)
244 are indeed out of the sensitive range of EPR measurements, making FRET a more suitable technique in
245 this context.

246     We note that the finite size of probes ($r_p \sim 1.7$ nm) brings additional difficulties to converting FRET
247 measurements to estimation of distances: FRET results gave the distances between the chromophores of
248 donors and acceptors, which is different from the distances between the $C_\alpha$ atoms of residues of interest.
249 However, on the other hand, the *movement* of the residues (or the movement of the $C_\alpha$ atoms of the
250 residues) in the radial direction is the same as the *movement* of the chromophores assuming that the size
251 of the probes does not change (i.e. $\Delta r_p = 0$) upon channel opening (see Supplementary Information for
252 details). We also note that, although the five-fold symmetry is broken due to the binding of one donor and
253 one acceptor per pentamer, the geometric construction will not be affected.

254     As a result, we focus on the more relevant distance of interest: the movement of the residue away
255 from the pore center, $\Delta r$ (Figure 5b), or the change of protein diameter measured from the residue, $\Delta D$. ($D$
256 is the protein diameter defined by a specific residue, as shown in Figure 5a-b). Because of the five-fold
257 symmetry of the MscL channel, $\Delta D$ and $\Delta r$ can be calculated readily according to $\Delta D = \Delta R_n / \sin(\pi/5) \approx$
258 2.8 nm, which yields $\Delta r = \Delta D / 2 \approx 1.4$ nm (for M42C). The $\Delta r$ values of the residues are summarized in
259 Table I. These values are all above 2.5 nm, a lower bound predicted by EPR experiments (Perozo *et al*,
260 2002a), but are larger than $\Delta D$ values obtained from the previous ensemble FRET measurement: for
261 example, $\Delta D_{M42C} = 2.8$ nm (smFRET) *vs.* $\Delta D_{M42C} = 1.6$ nm (ensemble FRET) (Corry *et al*, 2010, 2005b).
262 We emphasize that the measurements of two more residues (I25C and A27C) in the current study were
263 also reported previously (Corry *et al*, 2010). Our results are close to the values in their simulations
264 ($\Delta D_{I25C} = 2.4$ *vs.* 2.5 nm; $\Delta D_{A27C} = 2.5$ *vs.* 2.6 nm) but differ significantly from the values measured
265 directly from ensemble FRET experiments ($\Delta D_{I25C} = 2.4$ *vs.* 0.2 nm). It should be noted that ensemble
266 experiments gave inconsistent measurements for $\Delta D_{I25C} = 0.2$ nm and $\Delta D_{A27C} = 2.9$ nm, although the two
267 residues are close. In contrast, smFRET results show that $\Delta D_{I25C} = 2.4$ nm is similar to $\Delta D_{A27C} = 2.5$ nm.
268 This clearly demonstrates the advantage of smFRET.

269     We note that fluorophores/linkers at different residues are likely to be constrained differently.
270 Furthermore, how they are constrained differently is not clear, partly due to the unavailability of the
271 crystal structure of EcoMscL. However, certain residues are in agreement between the EcoMscL and the
272 MtMscL (Perozo *et al*, 2001). Nevertheless, the distances between donors and acceptors are *not* good to
273 compare for different residues of EcoMscL. A more reasonable way is to compare the *changes* of
274 distances, i.e., the movements of residues.

275     The calculations above were performed with the assumption that EcoMscL are pentamers. However, a
276 caveat is that, in certain detergents, a small fraction of EcoMscL proteins present as hexamers, instead of
277 pentamers (Gandhi *et al*, 2011). To estimate the uncertainties due to a mixture of pentamers and hexamers,
278 we performed quantitative numerical simulations and showed that our results would be smaller than the
279 actual values by 7.5% in the presence of 30% hexamers in the sample (Supplementary Figure 1 and 4).

280     Because the size of both Alexa fluorophores is significant ($\sim$1.7 nm), it is possible that the attachment
281 of the fluorophores to MscL channel results in various effects on the protein and on the FRET
282 measurements. For example, the presence of the fluorophores might sterically hinder the conformational
283 change of the proteins and prevent them from opening or closing. On the other hand, the steric hindrance
284 might constrain the orientation of fluorophores, affect the relative orientation between the fluorophores
285 and therefore add more errors on the distances converted from FRET efficiencies. In addition, the
286 insertion of fluorophores to the protein might force the channel to be in a state different from the fully

287 closed state, resulting in the distance change measururement is underestimated. However, we would like
288 to emphasize that the expected effect is insignificant for the following reasons. First, if the insertion of
289 fluorophore would result in significant steric hindrance on the protein, it is expected that the labeling is
290 difficult (i.e., it takes much more effort for the fluorophores to be attached due to the steric hindrance). In
291 other words, it is expected that steric hindrance is not significant on the mutants that are labeled well.
292 More importantly, the channels after being labeled with AF488 and AF568 were confirmed to be
293 functional by both ensemble FRET experiments (by observing the shift in the FRET peak) and patch-
294 clamp measurements (by observing the opening of the channels upon application of negative pressure to
295 the patch pipette) as shown in Figure 6 and previous publications with the same fluorophores (Corry et al,
296 2010).
297

**Computational MscL opening model.** With smFRET, we measured the movements of three residues on
299 TM1 (M42C, A27C, and I25C) and three on TM2 (Y75C, Q80C and V82C) summarized in Table I and
300 Figure 5c-d. We observed directly and reliably for the first time, that both TM1 and TM2 swing away
301 from the pore, supporting the helix-tilt model. Note that, among the three residues on each helix, two sites
302 were very close to each other (A27C and I25C on TM1, Q80C and V82C on TM2). They were chosen
303 purposefully to be close; they served as consistency checks and confirmed that our smFRET
304 measurements are accurate (Table I). In addition, the top of both helices (periplasmic side, Figure 1b-c;
305 residues 42 on TM1 and 75 on TM2) moves further than the bottom (1.4 nm vs. 1.2 nm for TM1 and 2.0
306 nm *vs.* 1.4 nm for TM2), indicating that rotational tilting of the helices (toward the membrane plane) is
307 involved. We emphasize that it is the first *direct* (model-free) observation of both TM1 and TM2
308 swinging away from the pore center and of the tilting of the transmembrane helices. Therefore it is the
309 first *direct* observation in favor of the helix-tilt model.
310    To quantitatively investigate in detail how the MscL channel opens (i.e. how the helices move and
311 rotate upon opening), we developed a computational model for the open structure of the MscL, starting
312 from the crystal structure of MscL in the closed state (PDB: 2OAR) (Chang *et al*, 1998; Steinbacher *et al*,
313 2007) and employing the measured residue movements. For this purpose, we performed MD simulation
314 with distance constraints (Brünger *et al*, 1986; Trabuco *et al*, 2009) (i.e., a virtual spring, Supplementary
315 Figure 5) using NAMD 2.9 (Phillips *et al*, 2005). Although similar modeling attempts have been made by
316 Corry et al. (Corry *et al*, 2010) and Deplazes et al. (Deplazes *et al*, 2012) by using distance changes
317 measured from ensemble FRET, we would like to emphasize that all smFRET measurements were used
318 for the simulation while previously only a selected subset of ensemble data were used (as other data were
319 not consistent with the resultant model) (Corry *et al*, 2010). For each measured residue, ten virtual springs
320 were placed, five springs between the central carbon atom $C_\alpha$ of identical residues (highlighted green in
321 Supplementary Figure 5) from adjacent monomers (red springs in Supplementary Figure 5) and five
322 springs between the $C_\alpha$ of identical residues from non-adjacent monomers (yellow springs in
323 Supplementary Figure 5). The virtual springs were not applied to side chains because the flexibility of
324 side chains likely introduces errors under large forces in the modeling process. The equilibrium lengths of
325 the springs were chosen by adding the distance changes measured from smFRET to the equilibrium
326 distances seen in the closed state, thereby, opening the crystal structure of *Mycobacterium tuberculosis*
327 MscL (PDB: 2OAR) (Steinbacher *et al*, 2007; Chang *et al*, 1998; Perozo *et al*, 2001). In the simulation,
328 the virtual springs pushed corresponding residues from the distance in the closed state to the equilibrium
329 length in the open state. We note that the uncertainty due to the size of the FRET probes was minimized
330 by focusing on the change of the distances between the closed and open state, rather than absolute
331 distances as discussed in previous section.
332    We note several limitations in the modeling: as the spring constant was kept constant through the
333 simulations, resulting in a large force at beginning of the simulation, we applied both secondary structure
334 restraints (Trabuco *et al*, 2009) and symmetry restraints (Chan *et al*, 2011) to prevent structural distortion.

The secondary structure restraints prevents some subtle changes in the structure, such as kinks observed previously in the upper part of TM1 in the open model of MscL (Deplazes *et al*, 2012). Therefore, we limit our discussion of the open model to pore size and helix tilting. The membrane tension, which causes membrane thinning, plays an important role in the MscL opening process (Corry *et al*, 2010; Louhivuori *et al*, 2010; Deplazes *et al*, 2012). However, the restraint MD simulation cannot address the question of how the channel is activated. For the simplicity of the modeling, membrane tension is not considered here. We did observe that the membrane near the MscL becomes thinner during the channel opening process to match with the flattening MscL (Supplementary Figure 6), confirming that a thinning membrane, likely caused by tension, matches the open channel better.

The resulting open state structure of MscL is shown in Figure. 7b and d, and compared with the crystal structure of MscL in the closed state (Figure 7a and c). The open structure satisfies all the distance constraints measured in our smFRET experiments. In contrast, previous models based on ensemble FRET measurements failed to be consistent with all experimental measurements (Corry *et al*, 2010). In the open conformation, the pore is mainly lined by helices TM1 (indicated by blue arrows), consistent with the helix-tilt model. In addition, it is observed that both TM1 and TM2 indeed tilt toward the membrane plane (horizontal) upon channel activation. For example, the orientation of TM1 tilts from the green arrow orientation (Figure 7c, closed state) to the yellow arrow orientation (Figure 7d, open state). The change in tiling angle of the TM1 and TM2 helices is $\Delta\theta_1 \approx 27°$ and $\Delta\theta_2 \approx 19°$, respectively, where $\theta$ is the angle between helix and the five-fold symmetry axis. The all-atom model and backbone model of the open state resulting from the current study are provided in PDB format in SI.

**Measurement of pore size in the open conformation.** We used two independent methods to measure the pore size of MscL in the open state. The first method is to measure the movements of the residues forming the narrowest pore constriction of the channel, i.e. residues around I25 for *E. coli* MscL (Chang *et al*, 1998; Perozo *et al*, 2001; Corry *et al*, 2010; Perozo *et al*, 2002a). However, this method, although straightforward, has its limitations. It is likely that the function of the channel is affected by mutation and labeling of (some of) the residues at the pore region. For example, the activation thresholds ($P_a$, defined as the pressure at which the first channel opening was observed (Nomura *et al*, 2012)) of mutants G22C and I24C are more than double the wild-type thresholds (Figure 6) and both ensemble and single molecule FRET measurements of these mutants showed no change in the FRET efficiency after adding 25% LPC. The effect of the point mutations near the pore on the electro-physiological properties of the channel can be quantitatively explained by the closed and open structure of MscL As shown in Supplementary Figure 10, the residue G22 (A20 in *Mycobacterium tuberculosis* MscL) is very close to the pore and is facing the pore. The residue V22 (V22 in *Mycobacterium tuberculosis* MscL) is also close to the pore and sandwiched between helix 1 and neighboring helix 1. Mutating these two residues is likely to perturb the channel function. On the other hand, the residue I25 is further from the pore than G22 and I24. The mutation I25C is less likely effect the channel properties. Indeed the I25C mutation does not affect the channel's gating parameters (Figure 6c-d). I25 is still close enough to the pore, making it a perfect candidate for measuring pore size. Furthermore, among the three mutated residues shown in Figure 6, I25 (green) is the only one facing outward from the channel axis and accessible from the periphery of the protein (see Supplementary Figure 10 B and D). We were able to determine the movement of residue I25C (Corry *et al*, 2010); and measured that the residue I25 moves away from the pore center by $\Delta r = 1.2$ nm, indicating that the pore opens up by $\Delta D = 2.4$ nm in diameter. Taking into account that the pore diameter in the closed state ($\Phi_{close}$) is 0.4 nm (Chang *et al*, 1998), we conclude that the pore size in the open state ($\Phi_{open}$) is $\Phi_{open} = \Phi_{close} + \Delta D = 2.8$ nm, which agrees with previously reported values (Perozo *et al*, 2002a; Corry *et al*, 2010).

The second method is based on the open state model of MscL constructed by means of molecular dynamics. The surfaces of water molecules inside the channel were rendered (Figure 7e-f) using VMD

383  (Humphrey *et al*, 1996) and the narrowest constriction seen provided an estimate of the pore size. This
384  estimate accounts for all residues of the transmembrane domain and therefore is expected to be more
385  accurate than the estimate of the first method. Using this method we estimate that the pore size of the
386  MscL channel in the fully open state is 2.7 – 2.8 nm, which is consistent with the value from the first
387  method, 2.8 nm.
388
389  **Discussion**
390  We used a combination of experimental smFRET and computational modeling to study the
391  conformational change of MscL upon channel activation. It is the first time that single molecule FRET
392  has been applied to liposome-reconstituted membrane proteins with more than three monomers. We
393  measured the distance changes of multiple residues from the MscL transmembrane α-helices (TM1 and
394  TM2) during gating of the channel. For the first time, it is observed *directly* that both transmembrane
395  helices swing away from the pore center, with rotational tilting involved. The results argue clearly in
396  favor of the helix-tilt model. In addition, we developed by means of computational modeling a model of
397  the channel structure in the open state based on the smFRET results and the crystal structure of the protein
398  in the closed state as a reference. This model also confirms the helix-tilt model and yields a pore diameter
399  of 2.8 nm. The smFRET experiments carried out in the present study observe MscL channels dynamics in
400  lipid bilayers (liposomes) and not in detergents, which is a great advantage over crystallography that can
401  result in different oligomeric states like those seen in the tetrameric structure of *S. aureus* MscL (Liu *et al*,
402  2009). It is possible that the detergent used in purification caused some portion ($\leq$ 30%) of the MscL as
403  hexamers, instead of the assume pentamers. Nevertheless, our conclusion of the helix-tilt opening model
404  is independent of the percentage of hexameric structure. However, the exact value of the open pore
405  diameter would be slightly greater, 3.0 nm (30% hexamers), up to 3.3 nm (100% hexamers), still agreeing
406  with previously reported values (Perozo *et al*, 2002a; Corry *et al*, 2010; Sukharev *et al*, 2001b).
407      The current study focused on the closed and fully open state of MscL. The fully open state was
408  achieved by adding LPC to the liposomes (Perozo *et al*, 2002a, 2002b). However, the technique
409  introduced is not limited to these two states only. Single molecule FRET together with other techniques--
410  for example, with patch-clamping done simultaneously--can answer many more questions than a crystal
411  structure. For instance, it could probe the conformation of the channel during sub-conducting levels that
412  involve partial MscL openings, or probe sequence of movements of the individual channel domains
413  during opening of the channel.
414
415  **Materials and methods**
416
417  **Mutation, expression, purification and labeling of MscL.** The *E. coli* MscL gene (EcoMscL) was
418  cloned into plasmid pQE-32 (Qiagen) as the *Bam*HI-*Sal*I fragment, which also added a hexa-histidine tag
419  (his-tag) to the protein at the N-terminus. The protein was expressed in *E. coli* (M15 strain) (Qiagen) that
420  were lysed by sonication and purified from DDM solubilized membranes using TALON® Metal affinity
421  chromatography (Clontech Laboratories, Inc), followed by a further purification step using fast protein
422  liquid chromatography (FPLC; Superdex 200 10/300 GL column, GE Healthcare, Pittsburgh, US).
423  Purification was performed in the presence of 1 mM DDM.
424      The wild type of MscL protein does not contain any cysteine. To label the proteins with fluorescent
425  probes, MscL was mutated using site-directed mutagenesis such that a residue at the desired position was
426  replaced by a cysteine. Because the MscL protein is a homo-pentamer (Chang *et al*, 1998), this mutation
427  introduced five identical cysteine sites.
428      The protein with his-tag was then labeled with Alexa Fluor 488 (AF488) and/or Alexa Fluor 568
429  (AF568) maleimide, which specifically reacted with the introduced cysteines (Kim *et al*, 2008). Right
430  before labeling, proteins were reduced with 10 mM DTT for 30 minutes, followed by purification using

PD-10 desalting columns (GE Healthcare, Pittsburgh, US). We titrated the pentameric protein-to-fluorophore molar ratio from 1:1 to 1:5 and used the molar ratio of 1:5 for labeling in all the experiments. Under our labeling conditions, this ratio gave satisfying results such that most of the proteins are labeled (averagely ~ 1.7 donors and ~ 1.3 acceptors per pentamer) and that many of proteins are attached by a single donor and a single acceptor (~ 30% of good traces show multiple donors and/or acceptors). Excess fluorophores were then removed using PD-10 desalting columns. The sample was reduced with 10 mM DTT before this purification step. A note to make is that the fluorophores (Alexa Fluor 488 maleimide and Alexa Fluor 568 maleimide) come as mixtures of 5' and 6' isomers, which would potentially complicate interpretation of smFRET data. However, we expect that the results would not be affected because the exactly same fluorophores have been successfully used in many single molecule FRET studies (Granier *et al*, 2007; Jäger *et al*, 2006; Majumdar *et al*, 2007; Marras *et al*, 2002; Yin *et al*, 2005).

**Reconstitution and opening of MscL in liposomes.** MscL channels were reconstituted into artificial liposomes (~ 50 nm diameter), following the protocol described in Ref. (Perozo *et al*, 2002b, 2002a). Liposomes were prepared by drying, rehydrating and extruding lipids through filters with ~ 50 nm pores. The lipids used in all the measurements were a mixture of 1-palmitoyl-2-oleoyl-sn-glycero-3-phosphocholine (POPC, Avanti polar lipids, Inc.) and 1,2-dioleoyl-sn-glycero-3-phosphoethanolamine-N-biotinyl (BPE, Avanti polar lipids, Inc.) dissolved in chloroform at a molar ratio of POPC:BPE = 1000:20. BPE was used for immobilization (see below). To incorporate MscL channels into the liposomes, a mixture of unlabeled and labeled MscL proteins (5% labeled) was then reconstituted into the liposomes, at a final volume of 1 ml, with a protein/lipid (molar) ratio of 1:200, resulting in a molar ratio of 1:4000 for the labeled proteins to lipids. The liposomes were immobilized onto a glass coverslip. This immobilization was achieved by biotin-avidin linkages between biotinylated-PEG molecules on the surface to a neutravidin molecule, and then biotinylated lipids (BPE) in the liposomes (Roy *et al*, 2008).

To open the MscL channels in the liposomes, a conical lipid, 1-oleoyl-2-hydroxy-sn-glycero-3-phosphocholine or lysophosphatidylcholine (LPC), was added to the liposomes, at a molar fraction of 25%. As LPC incorporates itself into the outer leaflet of a lipid bilayer, it introduces membrane tension, changes the lipid pressure profile, and triggers the MscL to open (Perozo *et al*, 2002b, 2002a).

**Electrophysiological recording.** MscL protein purification and reconstitution into soybean azolectin liposomes were described previously (Nomura *et al*, 2012). All results were obtained with proteoliposomes at the protein: lipid ratio of 1:200 (w/w). Channel activities of the wild-type and mutant MscL were examined in inside-out liposome patches using patch-clamp technique. Borosilicate glass pipettes (Drammond Scientific Co, Broomall, PA) were pulled using a Narishige micropipette puller (PP-83; Narishige, Tokyo, Japan). Pipettes with resistance of 2.5-4.9 M$\Omega$ were used for the patch-clamp experiments. Pipette and bath solution contained 200 mM KCl, 40 mM MgCl2, and 5 mM HEPES (pH 7.2 adjusted with KOH). The current was amplified with an Axopatch 200B amplifier (Molecular Devices, Sunnyvale, CA), filtered at 2 kHz and data acquired at 5 kHz with a Digidata 1440A interface using pCLAMP 10 acquisition software (Molecular Devices, Sunnyvale, CA) and stored for analysis. Negative pressure (suction) was applied to the patch pipettes using a syringe and was monitored with a pressure gauge (PM 015R, World Precision Instruments, Sarasota, FL).

**Selection of MscL with a single donor and a single acceptor.** Since the MscL channel is a homo-pentamer (Chang *et al*, 1998) (or possibly homo-hexamer (Gandhi *et al*, 2011)), there is always a distribution of various donor/acceptor combinations. To exclude signal from those channels having multiple donors or multiple acceptors, the fluorescence intensity of single channels (and hence the step-wise photobleaching) was monitored. Because multiple donors or acceptors have multiple "staircase" photobleaching, these channels were simply not used. Only the traces with a clear single-step photobleaching in both donor and acceptor channels were included in the analysis. Subtraction of the intensities (averaged) before and after photobleaching gives the intensities of donor ($I_D$) and acceptor ($I_A$), which are then used for FRET efficiency calculation as described below.

**Single molecule FRET measurement.** Single molecule FRET experiments were performed using total internal reflection fluorescence microscopy (TIRFM) with a 1.45 NA 100X oil immersion objective (Roy *et al*, 2008; Selvin & Taekjip, 2007). The fluorescence intensities were used to calculate the energy transfer efficiency by the corrected FRET equation: $E = (I_A − \ell\ I_D)/ (I_A + \gamma\ I_D)$: where E is the FRET efficiency, $\ell$ represents leakage of donor signals in the acceptor channel, $\gamma$ is the correction factor which accounts for the differences in quantum yield and detection efficiency between the donor and the acceptor, $I_A$ and $I_D$ represent the acceptor and donor intensities, respectively (Roy *et al*, 2008). Note that the direct excitation of the acceptor by the donor excitation has been corrected automatically when getting the acceptor intensity from the fluorescence traces. The distance between the donor and acceptor is given by $R = R_0(E^{-1}-1)^{1/6}$, where $R_0$ is the Förster radius (Förster, 1948). The Förster radius, $R_0$, given by $R_0 = \left(\frac{0.529\ \kappa^2\ Q_D\ J(\lambda)}{N_A\ n^4}\right)^{1/6} \propto (\kappa^2\ Q_D)^{1/6}$, and its error were measured experimentally by measuring the absorbance and fluorescence spectra, quantum yield of the donor, AF488, ($Q_D = Q_{AF488}$) and anisotropy ($A_a$ and $A_d$ which give the maximum possible error in $\kappa^2$) of the fluorescent probes conjugated to proteins.

**Measurement of quantum yield of AF488 conjugated to MscL.** The quantum yield of AF488 conjugated to MscL was measured using fluorescein in 0.1 M NaOH as a standard (Fery-Forgues & Lavabre, 1999; Lakowicz, 1999) using the equation $Q_X = \frac{A_S}{A_X} \times \frac{F_X}{F_S} \times \left(\frac{n_X}{n_S}\right)^2 \times Q_S$, where Q is the quantum yield, A is the absorbance at the excitation wavelength (470 nm); F is the area under the corrected emission curve, and n is the refractive index of the solvent. Subscripts S and X refer to the standard (fluorescein) and to the unknown (AF488), respectively. The spectra of absorbance and fluorescence of AF488-MscL in PBS+DDM (1mM DDM) were measured using Agilent 8453 UV-Vis absorbance spectrophotometer (Agilent technologies) and PC1 spectrofluorimeter (ISS, Inc.), respectively.

**Measurement of anisotropy of fluorophores conjugated to MscL.** In order to determine the maximum error in the orientation factor, $\kappa^2$, and therefore the error in $R_0$, the anisotropy of the fluorophores conjugated to MscL was measured. The fluorophores-protein conjugates were immobilized on a glass coverslip which was covered with PEG (5% biotinylated), then a layer of neutravidin (Thermo Scientific), followed by a layer of penta-his biotin conjugate (Qiagen). The emission of the fluorophores-protein conjugates were split into two channels of polarization and used to calculate the anisotropy, $A = \frac{I_\parallel - I_\perp}{I_\parallel + 2\ I_\perp}$, where $I_\parallel$ is the fluorescence emission with polarization parallel to the excitation polarization and $I_\perp$ is the fluorescence emission with polarization perpendicular to the excitation polarization (Lakowicz, 1999). Anisotropies were corrected for the intrinsic polarization properties of the microscope by calibrating to known freely diffusing fluorophores. Anisotropies were also corrected for the high numerical aperture of the objective. Then the maximum range of $\kappa^2$ was given by $\kappa^2_{max} = 2/3(1+2.5A_d+2.5A_a)$ and $\kappa^2_{min} = 2/3(1-1.25A_d-1.25A_a)$ where $A_d$ and $A_a$ are the anisotropy of AF488 (donor) and AF568 (acceptor), respectively (Cha *et al*, 1999; Dale *et al*, 1979).

**Modeling the MscL open structure through restraint molecular dynamics (MD) simulation.** Due to lack of an *E. coli* MscL (EcoMscL) crystal structure, the simulation were performed using the structure of MscL from *Mycobacterium tuberculosis* (MtMscL, PDB: 2OAR) (Steinbacher *et al*, 2007; Chang *et al*, 1998). The CP domain was truncated in the simulation because the complete deletion of the CP does not change the gating parameters substantially (Anishkin *et al*, 2003). The residues to which the distance constraints were applied, were shifted according to the sequence alignment in Ref. (Chang *et al*, 1998). A spring constant of 0.2 kcal mol$^{-1}$Å$^{-2}$ was used for the virtual spring in the distance constrained simulation. Both secondary structure restraints (Trabuco *et al*, 2009) and symmetry restraints (Chan *et al*, 2011) were applied to prevent structural distortion under large force in the distance constrained simulation. Total simulation time is 5 ns. A model of MscL in the open state was obtained at the end of the distance constrained simulation, when the simulation satisfied all the distance constraints measured by means of smFRET experiment. The restraint MD simulation procedure is similar to the one used previously (Corry *et al*, 2010; Deplazes *et al*, 2012).

The simulation system was prepared by first imbedding the crystal structure of MscL (PDB: 2OAR) (Steinbacher *et al*, 2007; Chang *et al*, 1998) into a membrane patch with 1727 POPC lipids. Solvent was then added to both sides of the membrane, and the system was neutralized with 200 mM NaCl using VMD (Humphrey *et al*, 1996). The final simulation system contained 1,137,413 atoms. The all-atom MD simulations were performed using NAMD (Phillips *et al*, 2005) with the TIP3P model (Jorgensen *et al*, 1983) for explicit water and the CHARMM36 force field (Best *et al*, 2012). The simulation was conducted in the NPT ensemble (constant pressure and temperature) with periodic boundary condition. Constant temperature of 300 K was maintained using a Langevin thermostat with a damping coefficient of 1 ps$^{-1}$. A Nosé–Hoover Langevin piston barostat was used to maintain a constant pressure of 1 atm with a period of 200.0 fs and damping timescale of 100.0 fs. The multiple time-stepping algorithm was employed, with an integration time step of 2 fs, the short-range force being evaluated every time step, and the long-range electrostatics every second time step. Non-bonded energies were calculated using particle mesh Ewald full electrostatics and a smooth (10–12 Å) cutoff of the van der Waals energy.

**Acknowledgements**

This work was supported by NIH Grants R01 GM068625, R01 GM067887, U54 GM087519, 9P41GM10460, and the NSF Grants PHY0822613 and OCI-1053575, and NH&MRC Grant 635525. The authors acknowledge supercomputer time on Stempede provided by the Texas Advanced Computing Center (TACC) at The University of Texas at Austin through Extreme Science and Engineering Discovery Environment (XSEDE) Grant MCA93S028. We thank Kai Wen Tseng for assistance with the quantum yield measurements. We also thank Eduardo Perozo (U. of Chicago) for early work and for some MscL plasmids.

**Author Contribution**

Y.W. and H.A.D. expressed and purified proteins. Y.W. performed single molecule FRET measurements. Y.L. and K.S. performed molecular dynamics simulations. T.N. and B.M. performed electrophysiological experiments. Y.W., Y.L., H.A.D. and T.N. analyzed data. M.T.H. and P.R.S. initiated the project. Y.W., Y.L. and P.R.S wrote the paper with the help of all the other authors.

**Competing Financial Interests Statement**

The authors declare no competing financial interests.

713
714
715  **Figures and Figure Legends**

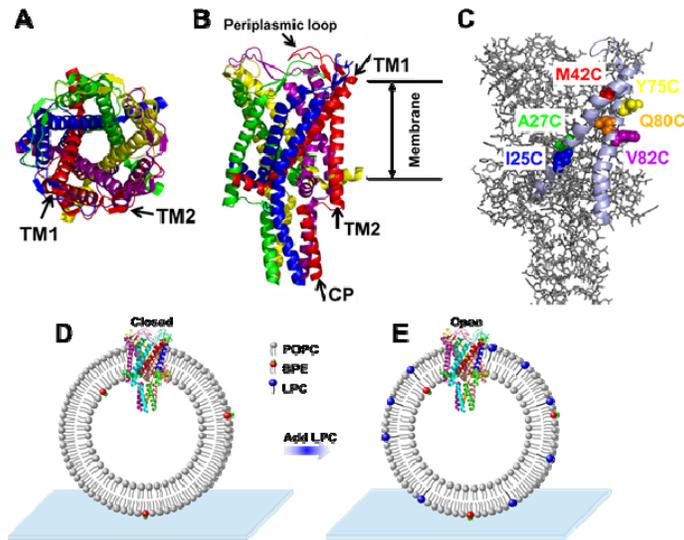

716
717  **Figure 1.  Cartoon representation of the structure of MscL in the closed conformation in the (a) top**
718  **view and (b) side view (PDB ID: 2OAR** (Chang *et al*, 1998; Steinbacher *et al*, 2007)**), and scheme of**
719  **single molecule FRET setup.** MscL is a homo-pentamer consisting of five identical subunits. Each
720  subunit consists of one cytoplasmic α-helix (CP) and two trans-membrane α-helices (TM1 and TM2),
721  which extend through the cell membrane and are joined by a periplasmic loop (Chang *et al*, 1998). (c)
722  Residues measured using smFRET. Three residues on each of the transmembrane helices (M42C, A27C
723  and I25C on TM1; Y75C, Q80C and V82C on TM2) were chosen. Note that no residues on the CP were
724  chosen because the complete deletion of the CP does not change the gating parameters substantially
725  (Anishkin *et al*, 2003). (d) Labeled MscL proteins were reconstituted into liposomes, which were then
726  immobilized on a coverslip and used for smFRET experiments. (e) The addition of LPC traps the protein
727  in the open conformation (Perozo *et al*, 2002b).

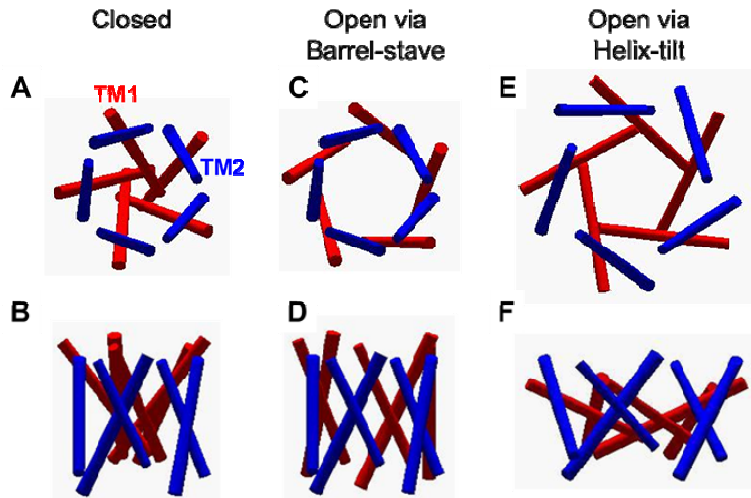

Figure 2. The opening models for MscL. The MscL opens from (a, b) the closed state, to (c, d) the open state via the barrel-stave model or (e, f) the open state via the helix-tilt model. The top figures (a, c, e) are top views and the bottom figures (b, d, f) are the side views. TM1 helices are shown in red while TM2 in blue. In the barrel-stave model (c, d), TM1 swings away from the pore center but TM2 remains stationary upon channel activation, generating an open pore lined by both TM1 and TM2 and the helices are more parallel to the membrane normal than the membrane plane. In the helix-tilt model (e, f), both TM1 and TM2 swing away from the symmetry axis and both helices tilt toward the plane of membrane.

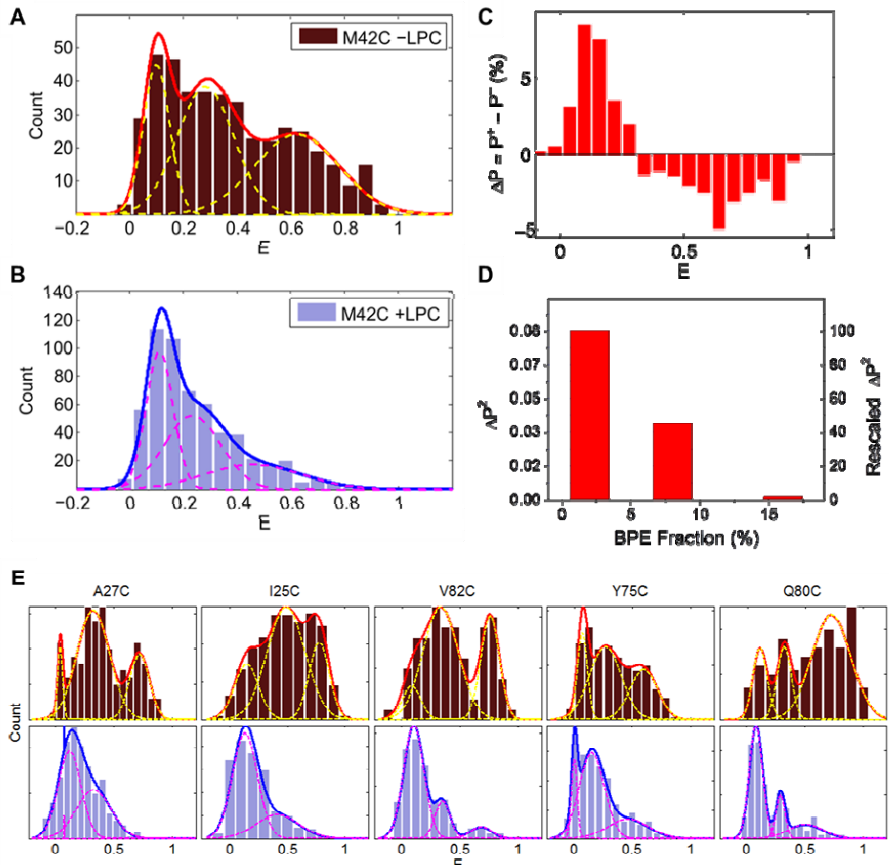

Figure 3. Single molecule FRET results for MscL M42C. The distribution of FRET efficiency of M42C in the (a) absence and (b) presence of LPC were plotted and fitted with Gaussians. (c) The difference between the normalized FRET distributions under the two conditions (±LPC), $\Delta P$, emphasizes

the diminishing of the third peak at E ~ 0.6 after adding LPC. (d) The variance between the normalized FRET distributions under the two conditions (±LPC), $\Delta P^2$, decreases as the fraction of BPE in the liposomes is increased from 2% to 16%. (e) Histograms of FRET efficiencies in the absence (top row, - LPC) and presence (bottom row, +LPC) of LPC for the other five residues (I25C, A27C, Y75C, Q80C, and V82C) that were measured in the current study.

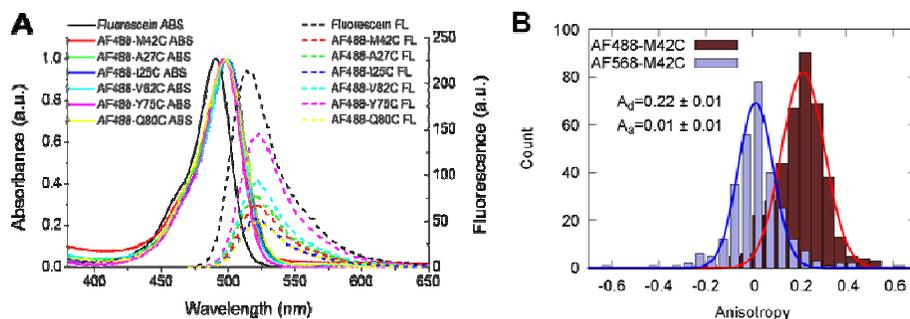

**Figure 4. Measurement of $R_0$.** (a) Absorbance and fluorescence spectra of AF488-MscL and fluorescein (as a standard), used to determine the quantum yield of AF488 conjugated to MscL mutants. (b) Anisotropy of AF488 and AF568 conjugated to MscL mutant (M42C), corrected for the intrinsic polarization properties of the microscope, and for the high numerical aperture of the objective.

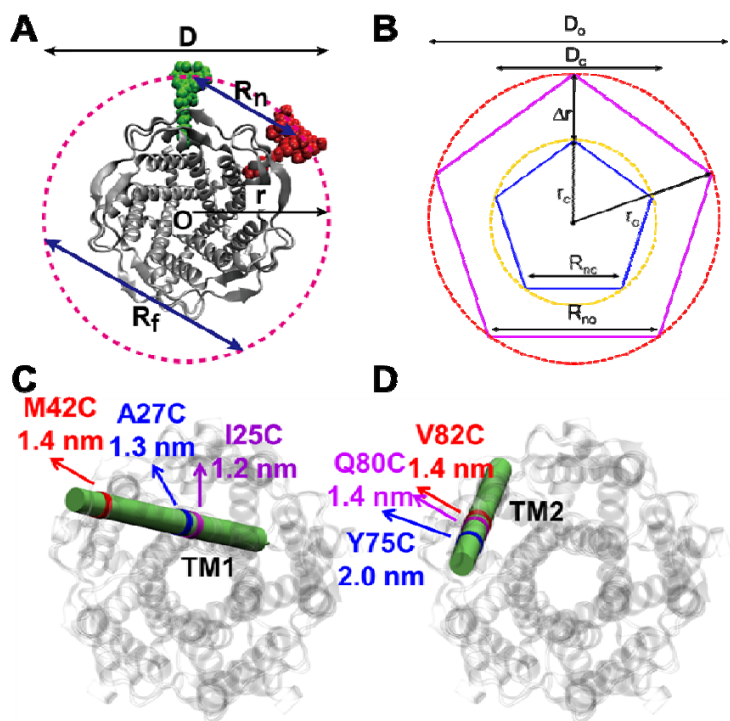

**Figure 5. Movement of residues.** (a) Each residue (highlighted in green) defines a circumcircle (dashed red circle) of radius r (or diameter D, where D, as shown, is $D_{closed}$, although upon opening would be $D_{open}$), centered at the pore center ($O$). Upon channel activation, the protein expands (radius changes from $r_{close}$ to $r_{open}$), or equivalently, the residue moves by $\Delta r = r_{open} - r_{close}$, measured from the pore center ($O$). (b) Sketch of MscL from closed state (blue pentagon) to open state (purple pentagon). The residue of interest (vertices of the pentagons) moves $\Delta r$ from the pore center. (c, d) Translational movements ($\Delta r$) of residues on TM1 and TM2 measured via smFRET. All the residues move away from the pore center, arguing in favor of the helix-tilt model.

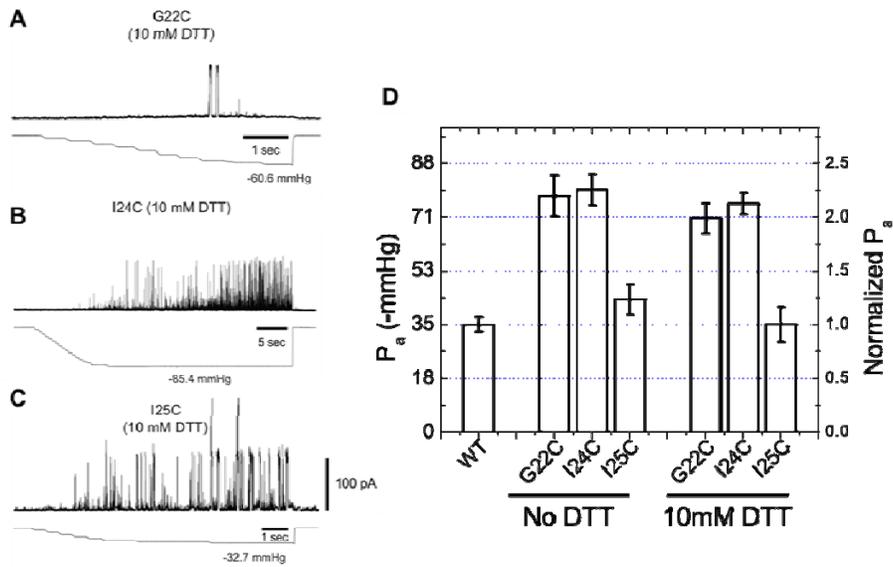

Figure 6. Activation thresholds, $P_a$, of MscL mutants at the proximity of the narrowest pore constriction. The activation thresholds were determined by electro-physiological recordings by patch-clamping without and with 10 mM DTT. Three recordings in the presence of DTT are shown as examples: (a) G22C, (b) I24C, and (c) I25C. (d) Comparison of the mutants with the wild type (WT) shows that the thresholds for mutants G22C and I24C are more than twice higher than the wild type, indicating the function of the channel was affected by the mutations. This was also observed via ensemble and single molecule FRET experiments. However, the mutation I25C does not affect the gating parameter substantially.

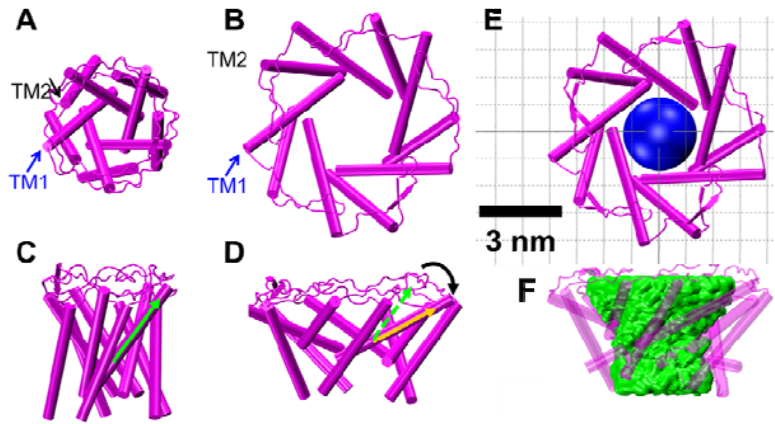

Figure 7. Model of the MscL structure in the open conformation. (a, c) The crystal structure of MscL in the closed state is shown for comparison (PDB: 2OAR (Chang *et al*, 1998; Steinbacher *et al*, 2007)). (b, d) The structure of MscL in the open state was developed based on the smFRET measurements, satisfying all the distance constraints measured from smFRET experiments. In the open conformation, the pore is mainly lined by TM1 (indicated by blue arrows), consistent with the helix-tilt model. In addition, both TM1 and TM2 tilt toward the membrane plane (horizontal) upon channel activation, which is emphasized by the green and yellow arrows in the side views. The green arrows show the orientation of TM1 in the closed state while the yellow arrow indicated the orientation of TM1 in the open state. The angle between the two arrows is 27°. (e) A sphere with a diameter of 2.7 nm (blue) is shown in the MscL channel in the top view. (f) The surfaces of water molecules (green) inside the tunnel of MscL (magenta) are drawn and the narrowest constriction is ~ 2.7 – 2.8 nm.

**Table I**. **Measurements of smFRET experiments.** $Q_d$ is the quantum yield of donor (AF488) after conjugation to each MscL mutant. $A_d$ and $A_a$ are the anisotropy of donor (AF488) and acceptor (AF568) after conjugation, respectively. $R_0$ is the Förster radius. $E_{nc}$ and $E_{no}$ are the FRET efficiencies in the closed and open states, respectively. $\Delta R$ is the change in the distances between donor and acceptor ($\Delta R_n = R_{no} - R_{nc}$, Figure 5a-b). $\Delta D$ is the change of the protein diameter ($\Delta D = D_{open} - D_{close}$). $\Delta r$ is the translational movement of the residue, measured from the pore center, $\Delta r = \Delta D / 2$. Note that the errors shown in the table are the maximum possible errors due to anisotropic orientation of the dyes. The actual errors are expected to be much smaller.

| Residue | Helix | $Q_d$ | $A_d$ | $A_a$ | $R_0$ (nm) | $E_{nc}$ | $E_{no}$ | $\Delta R_n$ (nm) | $\Delta r$ (nm) | $\Delta D$ (nm) |
|---------|-------|-------|-------|-------|------------|----------|----------|-------------------|-----------------|-----------------|
| 42 | TM1 | 0.33 | 0.22 | 0.01 | $5.5^{+0.4}_{-0.3}$ | 0.63 | 0.23 | $1.7^{+0.7}_{-0.5}$ | $1.4^{+0.6}_{-0.4}$ | $2.8^{+1.1}_{-0.8}$ |
| 27 | TM1 | 0.28 | 0.12 | 0.09 | $5.3^{+0.4}_{-0.3}$ | 0.72 | 0.33 | $1.5^{+0.6}_{-0.4}$ | $1.3^{+0.5}_{-0.3}$ | $2.5^{+0.9}_{-0.6}$ |
| 25 | TM1 | 0.42 | 0.19 | 0.06 | $5.7^{+0.5}_{-0.4}$ | 0.78 | 0.42 | $1.4^{+0.6}_{-0.5}$ | $1.2^{+0.6}_{-0.4}$ | $2.4^{+1.1}_{-0.8}$ |
| 75 | TM2 | 0.62 | 0.19 | 0.11 | $5.6^{+0.6}_{-0.5}$ | 0.60 | 0.16 | $2.4^{+1.0}_{-0.7}$ | $2.0^{+0.8}_{-0.6}$ | $4.0^{+1.6}_{-1.2}$ |
| 80 | TM2 | 0.22 | 0.18 | 0.09 | $5.1^{+0.5}_{-0.3}$ | 0.72 | 0.29 | $1.6^{+0.7}_{-0.4}$ | $1.4^{+0.6}_{-0.4}$ | $2.7^{+1.1}_{-0.8}$ |
| 82 | TM2 | 0.39 | 0.24 | 0.13 | $6.1^{+0.7}_{-0.6}$ | 0.76 | 0.35 | $1.6^{+0.9}_{-0.8}$ | $1.4^{+0.8}_{-0.6}$ | $2.7^{+1.5}_{-1.3}$ |

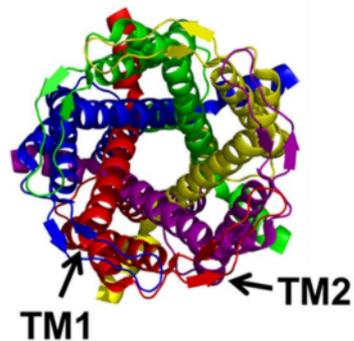

**A**

TM1

TM2

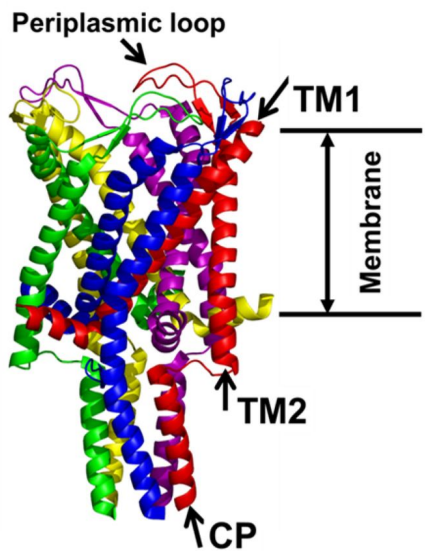

**B**

Periplasmic loop

TM1

Membrane

TM2

CP

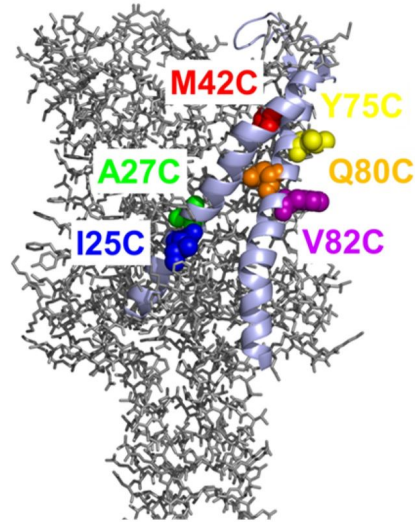

**C**

M42C

A27C

I25C

Y75C

Q80C

V82C

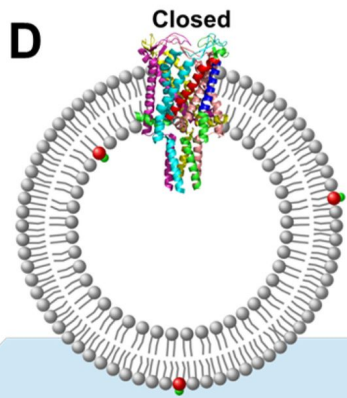

**D**

Closed

POPC

BPE

LPC

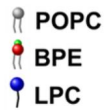

Add LPC

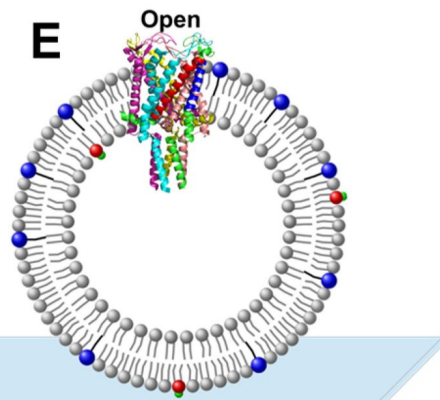

**E**

Open

**Closed**   **Open via Barrel-stave**   **Open via Helix-tilt**

**A**
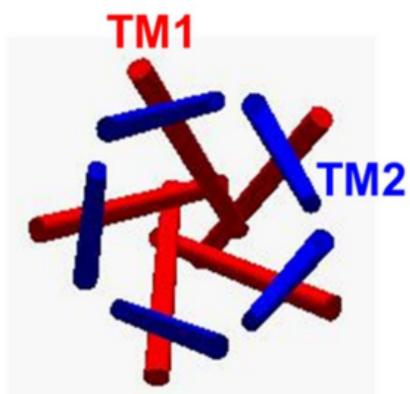
TM1
TM2

**C**
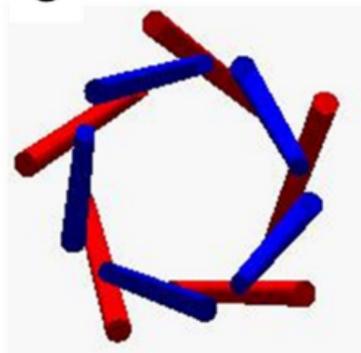

**E**
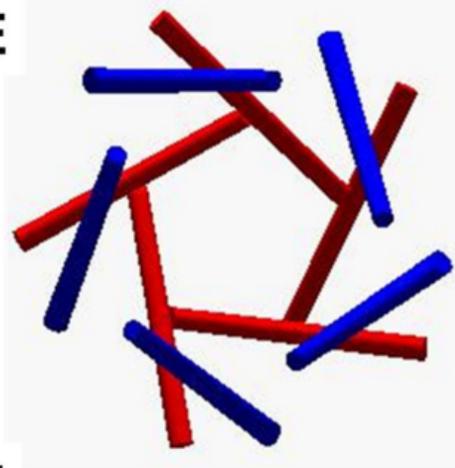

**B**
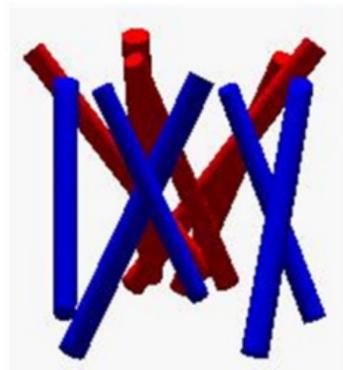

**D**
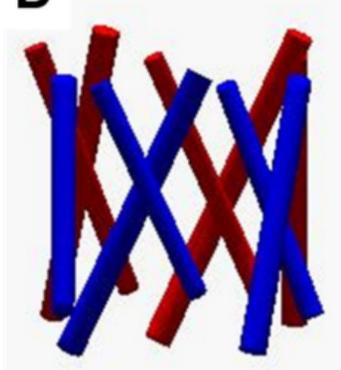

**F**
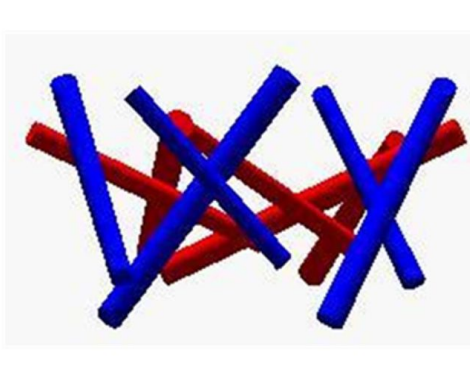

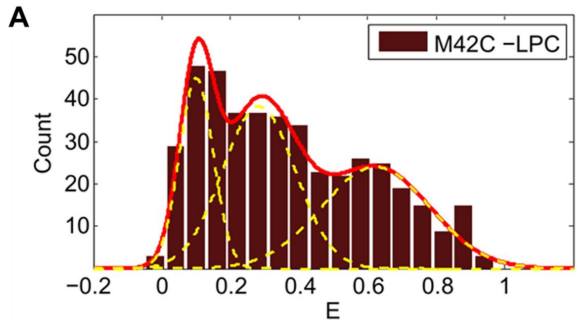

**A**

**B**

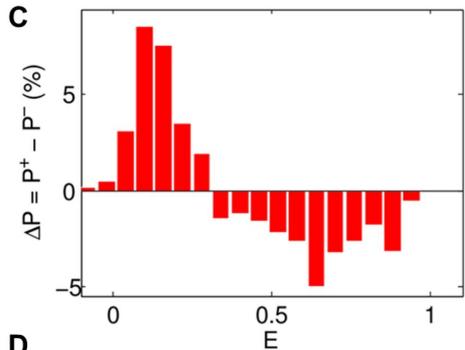

**C**

**D**

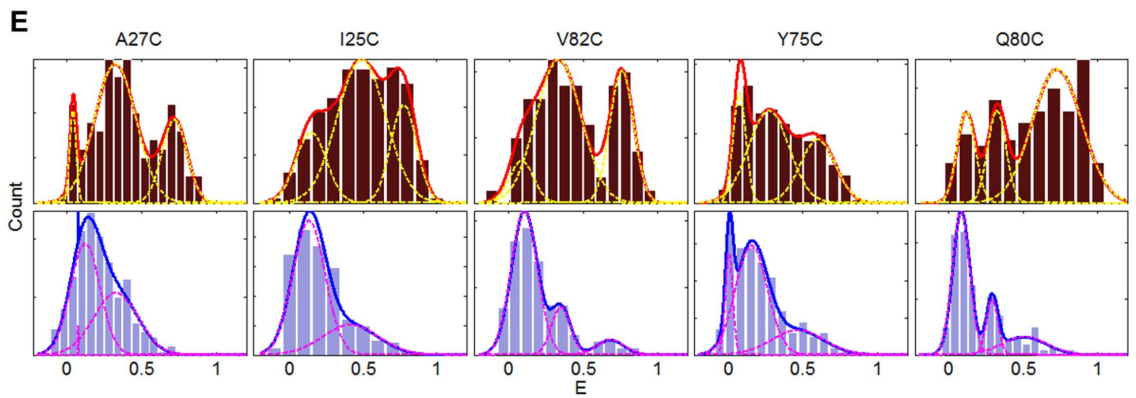

**E**

A27C  I25C  V82C  Y75C  Q80C

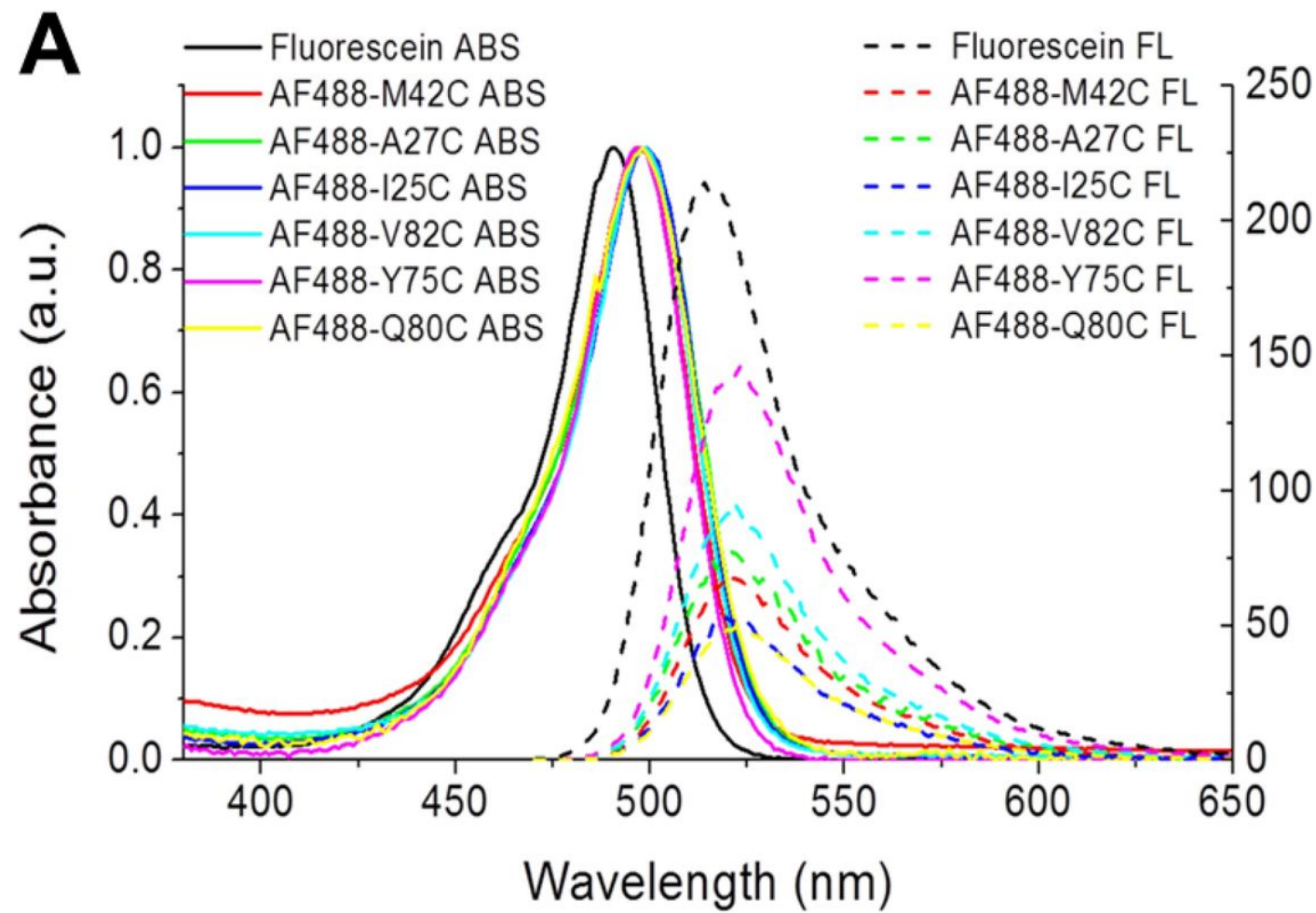 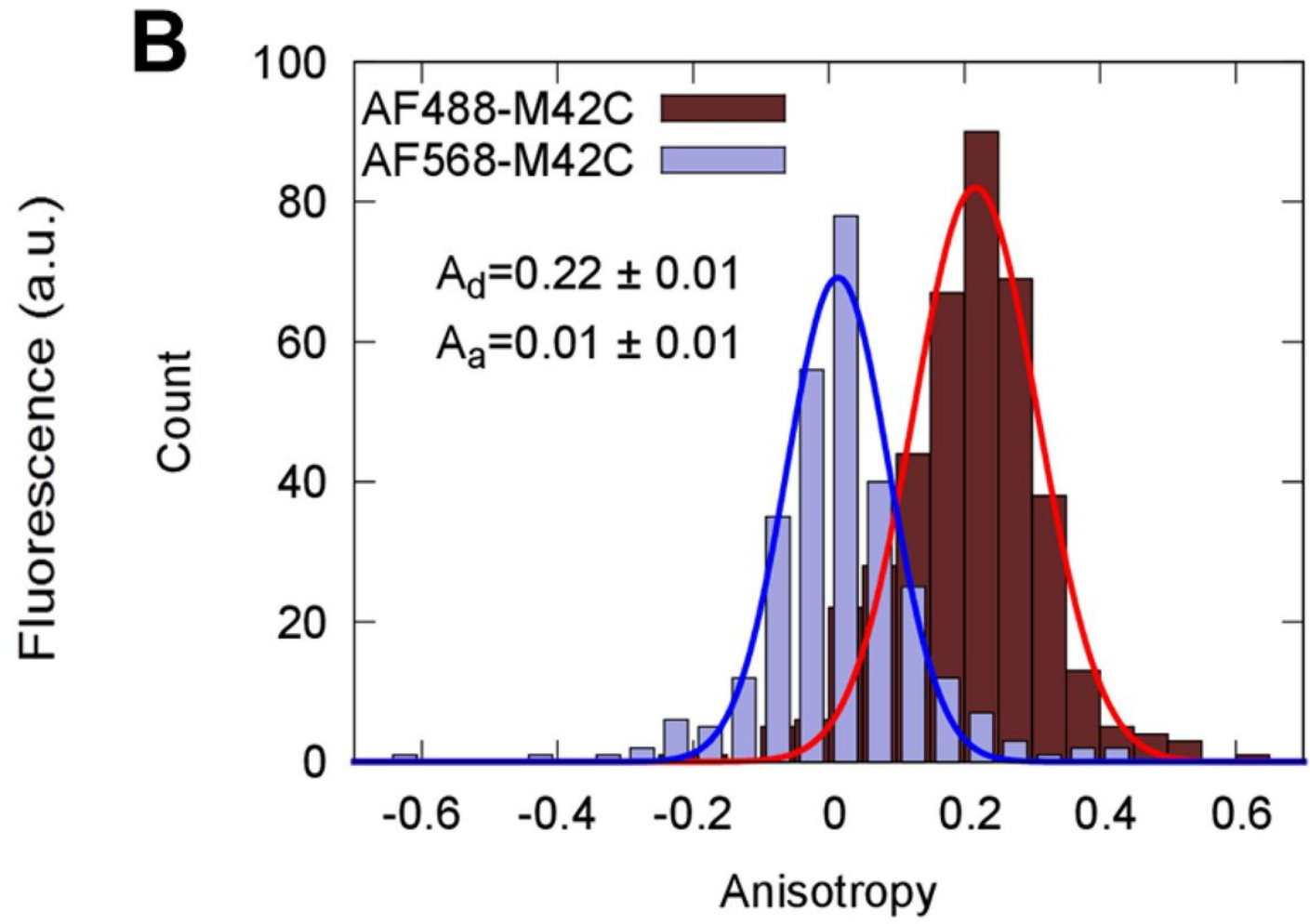

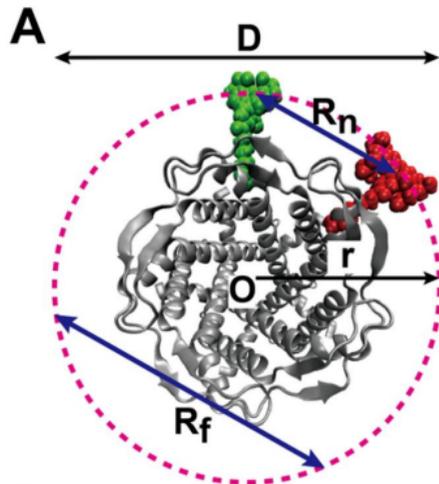

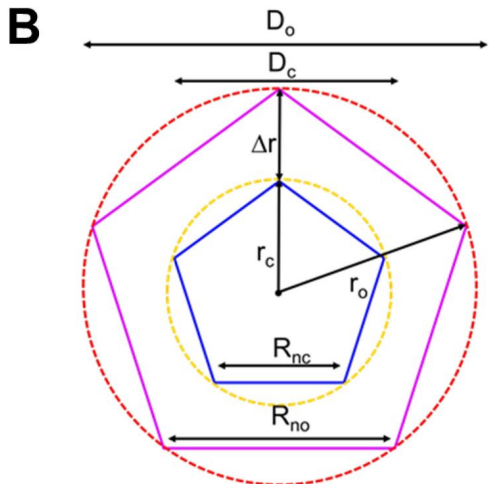

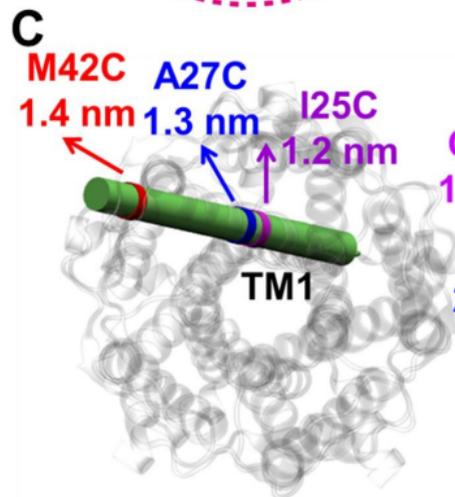

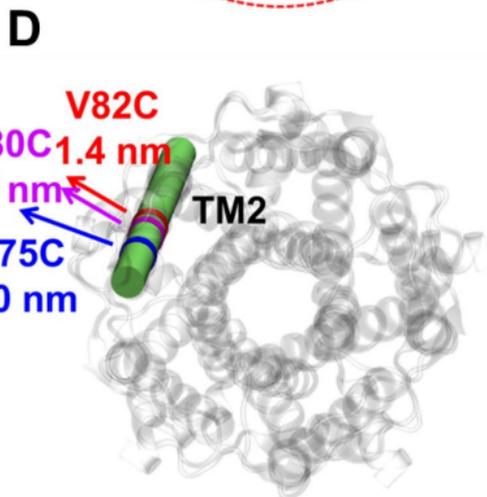

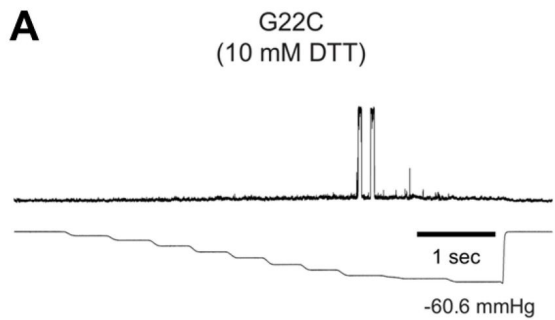

**A** G22C
(10 mM DTT)

1 sec

−60.6 mmHg

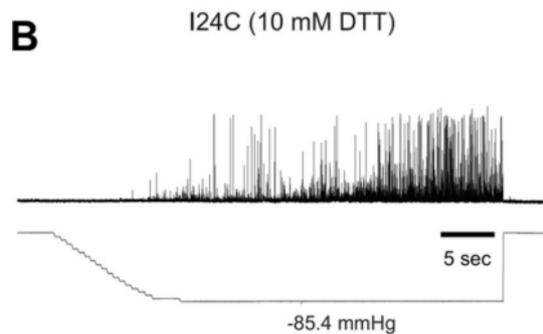

**B** I24C (10 mM DTT)

5 sec

−85.4 mmHg

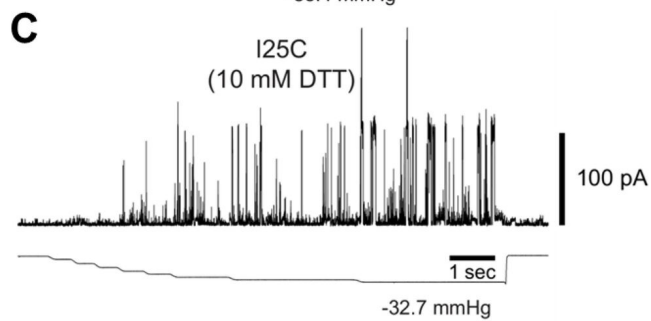

**C** I25C
(10 mM DTT)

100 pA

1 sec

−32.7 mmHg

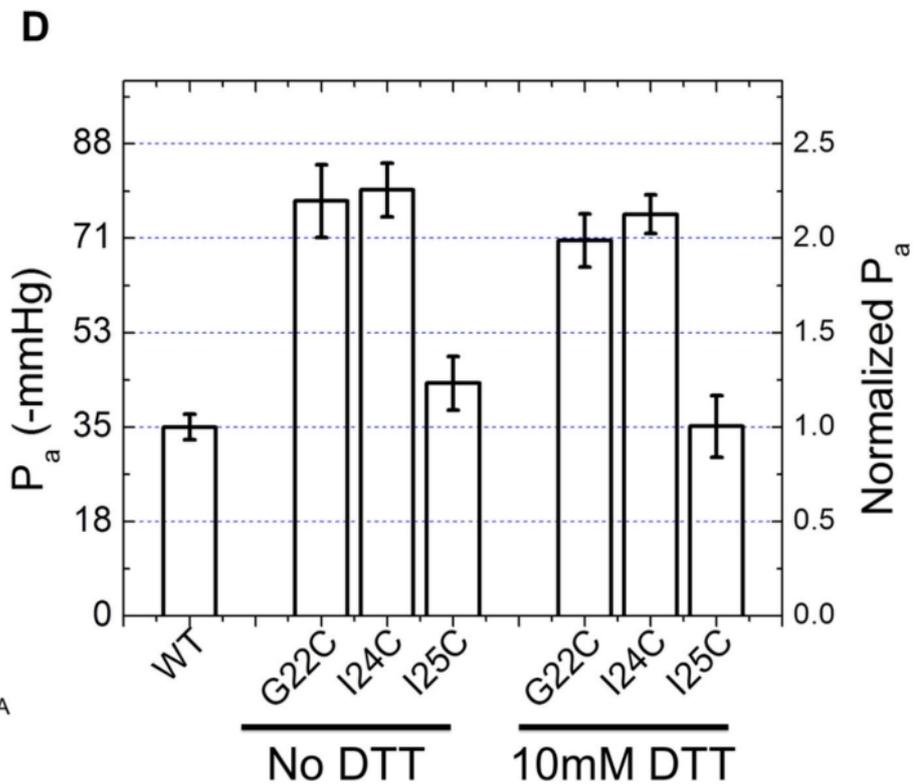

**D**

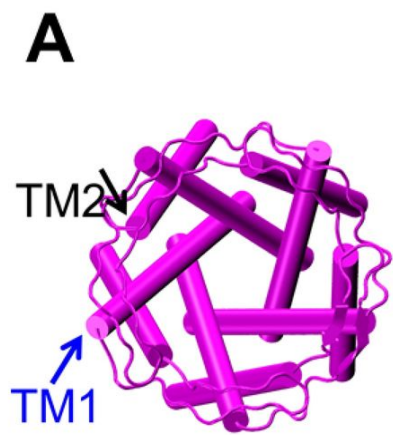

A

TM2

TM1

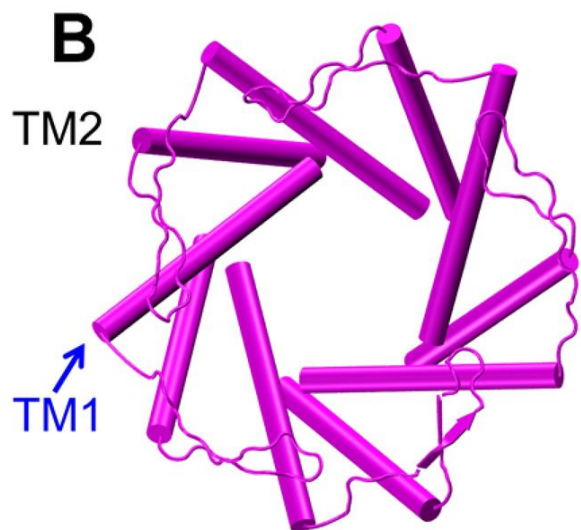

B

TM2

TM1

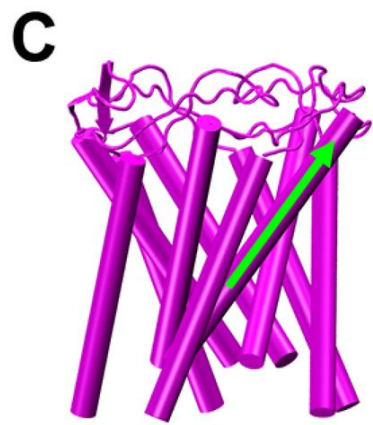

C

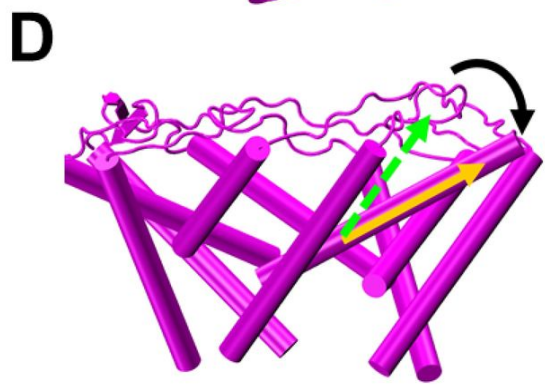

D

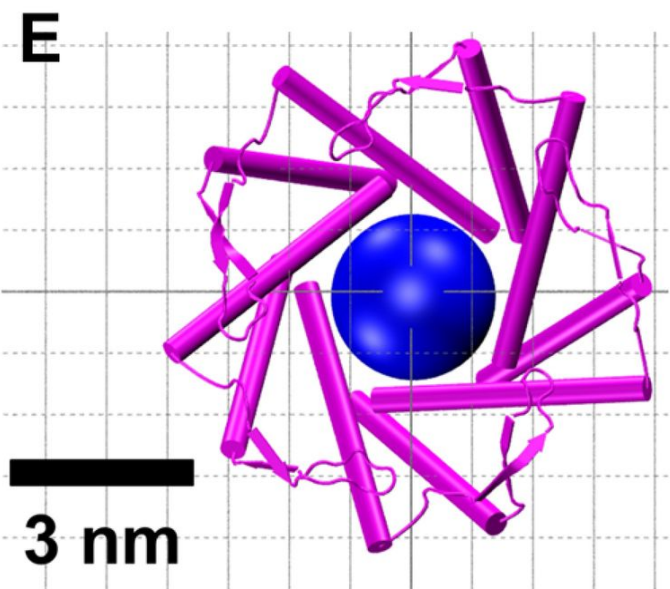

E

3 nm

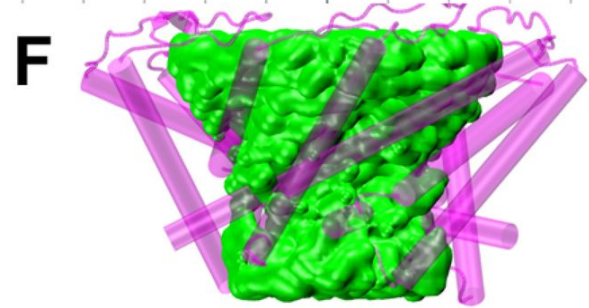

F